 \documentclass [12pt,a4paper      ]{article}
\usepackage{times}

\DeclareFontFamily{OT1}{times}{}
\DeclareFontShape {OT1}{times}{m }{n }{ <-> ptmr }{}
\DeclareFontShape {OT1}{times}{bx}{n }{ <-> ptmb }{}
\DeclareFontShape {OT1}{times}{m }{it}{ <-> ptmri}{}
\DeclareFontShape {OT1}{times}{bx}{it}{ <-> ptmbi}{}
\begin{document}

\title{\bf\vspace{-2.5cm} {\it From the Lab to the Battlefield?}\\Nanotechnology and Fourth Generation Nuclear Weapons\footnote{Published \emph{in} Disarmament Diplomacy, No.\,67 (October-November 2002) 3--6. Slightly expanded version with a few additional end-notes and references.}}

\author{{\bf Andre Gsponer}\\
\emph{Independent Scientific Research Institute}\\ 
\emph{ Box 30, CH-1211 Geneva-12, Switzerland}\\
e-mail: isri@vtx.ch}

\date{Version ISRI-02-06.10 ~~ \today}

\maketitle

\begin{abstract}

The paper addresses some major implications of microelectromechanical systems (MEMS) engineering and nanotechnology for the improvement of existing types of nuclear weapons, and the development of more robust versions of these weapons, as well as for the development of fourth generations nuclear weapons in which nanotechnology will play an essential role.

\end{abstract}

\section{Introduction}
\label{int:0}

\emph{In \emph{Disarmament Diplomacy} No.\,65, Sean Howard warned of the dangers of enhanced or even new types of weapons of mass destruction (WMD) emerging from the development of `nanotechnology', an umbrella term for a range of potentially revolutionary engineering techniques at the atomic and molecular level \cite{HOWAR2002-}.   Howard called for urgent preliminary consideration to be given to the benefits and practicalities of negotiating an `inner space treaty' to guard against such developments. While echoing this call, this paper draws attention to the \emph{existing} potential of nanotechnology to affect dangerous and destabilizing `refinements' to existing nuclear weapon designs. Historically, nanotechnology is a child of the nuclear weapons labs, a creation of the WMD-industrial complex. The most far-reaching and fateful impacts of nanotechnology, therefore, may lie --- and can already be seen --- in the same area.\\
\rule{0mm}{1mm} \hfill Sean Howard, Editor}

\section{The Strategic Context}
\label{str:0}

Two important strategic lessons were taught by the last three wars in which the full extent of Western military superiority was displayed: Iraq, Yugoslavia, and Afghanistan. First, the amount of conventional explosive that could be delivered by precision-guided munitions like cruise missiles is ridiculous in comparison to their cost: some targets could only be destroyed by the expenditure of numerous delivery systems while a single one loaded with a more powerful warhead would have been sufficient \cite{TELLE1991-}.  Second, the use of weapons producing a low level of radioactivity appears to be acceptable, both from a military point of view because such a level does not impair further military action, and from a political standpoint because most political leaders, and shapers of public opinion, did not object to the battlefield use of depleted uranium \cite{Note-DU}. 
These lessons imply a probable military perception of the need for new conventional or nuclear warheads, and a probable political acceptance of such warheads if they do not produce large amounts of residual radioactivity. Moreover, during and after these wars, it was often suggested that some new earth-penetrating weapon was needed to destroy deeply buried command posts, or facilities related to weapons of mass destruction \cite{BAILE1994-}.

It is not, therefore, surprising to witness the emergence of a well-funded scientific effort apt to create the technological basis for making powerful new weapons --- an effort that is not sold to the public opinion and political leaders as one of maintaining a high level of military superiority, but rather as one of extending human enterprise to the next frontier: the inner space of matter to be conquered by the science of nanotechnology.

\section{The Military Impact of Nanotechnology}
\label{mil:0}

Nanotechnology, i.e., the science of designing microscopic structures in which the materials and their relations are machined and controlled atom-by-atom, holds the promise of numerous applications. Lying at the crossroads of engineering, physics, chemistry, and biology, nanotechnology may have considerable impact in all areas of science and technology. However, it is certain that the most significant near term applications of nanotechnology will be in the military domain. In fact, it is under the names of `micromechanical engineering' and `microelectromechanical systems' (MEMS) that the field of nanotechnology was born a few decades ago --- in nuclear weapons laboratories.

A primary impetus for creating these systems was the need for extremely rugged and safe arming and triggering mechanisms for nuclear weapons such as atomic artillery shells. In such warheads, the nuclear explosive and its trigger undergo extreme acceleration (10,000 times greater than gravity when the munition is delivered by a heavy gun). A general design technique is then to make the trigger's crucial components as small as possible \cite{Note-size}.  For similar reasons of extreme safety, reliability, and resistance to external factors, the detonators and the various locking mechanisms of nuclear weapons were increasingly designed as more and more sophisticated microelectromechanical systems. Consequently, nuclear weapons laboratories such as the Sandia National Laboratory in the US are leading the world in translating the most advanced concepts of MEMS engineering into practice \cite{AMATO1998-}.

A second historical impetus for MEMS and nanotechnology, one which is also over thirty years old, is the still ongoing drive towards miniaturization of nuclear weapons and the related quest for very-low yield nuclear explosives which could also be used as a source of nuclear energy in the form of controlled microexplosions. Such explosions (with yields in the range of a few kilograms to a few tons of high-explosive equivalent) would in principle be contained – but they could just as well be used in weapons if suitable compact triggers are developed. In this line of research, it was soon discovered that it is easier to design a micro-fusion than a micro-fission explosive (which has the further advantage of producing much less radioactive fallout than a micro-fission device of the same yield). Since that time enormous progress has been made, and the research on these micro-fusion bombs has now become the main advanced weapons research activity of the nuclear weapons laboratories, using gigantic tools such as the US National Ignition Facility (NIF) and France's Laser M\'egajoule. The tiny pellets used in these experiments, containing the thermonuclear fuel to be exploded, are certainly among the most delicate and sophisticated nano-engineered devices in existence.

A third major impetus for nanotechnology is the growing demand for better materials (and parts made of them) with extremely well characterized specifications. These can be new materials such as improved insulators which will increase the storage capacity of capacitors used in detonators, nano-engineered high-explosives for advanced weaponry, etc. But they can also be conventional materials of extreme purity, or nano-engineered components of extreme precision. For instance, to meet NIF specifications, the 2-mm-diameter fuel pellets must not be more than 1 micrometer out of round; that is, the radius to the outer surface can vary by no more than 1 micrometer (out of 1,000) as one moves across the surface. Moreover, the walls of these pellets consist of layers whose thicknesses are measured in fractions of micrometers, and surface-smoothnesses in tens of nanometers; thus, these specifications can be given in units of 1,000 or 100 atoms, so that even minute defects have to be absent for the pellets to implode symmetrically when illuminated by the lasers.

The final major impetus for MEMS and nanotechnology, which has the greatest overlap with non-military needs, is their promise of numerous new very-high performance sensors, transducers, actuators, and electronic components. The development of this field of applications is expected to replicate that of the micro-electronic industry, which was also originally driven by military needs, and which provides the reference for forecasting a nano-industrial boom and a financial bonanza. There are, however, two major differences. First, electronic devices which can manufactured in large quantities and at low cost are essentially planar, while MEMS are three-dimensional devices which may include moving parts. Second, the need for MEMS outside professional circles (medical, scientific, police, military) is quite limited, so that the market might not be as wide as expected. For example, the detection and identification of chemical or biological weapon threats through specificity of molecular response may lead to all sorts of medical applications, but only to few consumers goods.

\section{Near- and Long-Term Applications and Implications of Nanotechnology}
\label{nea:0}

Considering that nanotechnology is already an integral part of the development of modern weapons, it is important to realize that its immediate potential to improve existing weapons (either conventional or nuclear), and its short-term potential to create new weapons (either conventional or nuclear), are more than sufficient to require the immediate attention of diplomats and arms controllers.
In this perspective, the potential long-term applications of nanotechnology (and their foreseeable social and political implications) should neither be downplayed nor overemphasized. Indeed, there are potential applications such as self-replicating nano-robots (`nanobots') which may never prove to be feasible because of fundamental physical or technical obstacles \cite{SMALL2001-}.  But this impossibility would not mean that the somewhat larger micro-robots of the type that are seriously considered in military laboratories could never become a reality \cite{nanobots}. 

In light of these extant and potential dangers and risks, every effort should be made not to repeat the error of the arms-control community with regard to missile defense. For over thirty years, that community acted on the premise that a ballistic missile defense system will never be built because it will never be sufficiently effective --- only to be faced with a concerted attempt to construct such a system! If some treaty is contemplated in order to control or prohibit the development of nanotechnology, it should be drafted in such a way that all reasonable long-term applications are covered. 

Moreover, it should not be forgotten that while nanotechnology mostly emphasizes the \emph{spatial} extension of matter at the scale of the nanometer (the size of a few atoms), the \emph{time} dimension of mechanical engineering has recently reached its ultimate limit at the scale of the femtosecond (the time taken by an electron to circle an atom). It has thus become possible to generate bursts of energy in suitably packaged pulses in space and time that have critical applications in nanotechnology, and to focus pulses of particle or laser beams with extremely short durations on a few micrometer down to a few nanometer sized targets. The invention of the `superlaser', which enabled such a feat and provided a factor of one million increase in the instantaneous power of tabletop lasers, is possibly the most significant recent advance in military technology. This increase is of the same magnitude as the factor of one million difference in energy density between chemical and nuclear energy \cite{Note-SL}.

In the present paper, the long-term impact of nanotechnology will not be further discussed. The objective is to emphasize the near to short-term applications to existing and new types of nuclear weapons.

\section{Nanotechnological Improvement of Existing Types of Nuclear Weapons}
\label{imp:0}

Nuclear weapon technology is characterized by two sharply contrasting demands. On the one hand, the nuclear package containing the fission and fusion materials is relatively simple and forgiving, i.e., rather more sophisticated than complicated. On the other hand, the many ancillary components required for arming the weapon, triggering the high-explosives, and initiating the neutron chain-reaction, are much more complicated. Moreover, the problems related to maintaining political control over the use of nuclear weapons, i.e., the operation of permissive action links (PALs), necessitated the development of protection systems that are meant to remain active all the way to the target, meaning that all these ancillary components and systems are submitted to very stringent requirements for security, safety, and reliable performance under severe conditions.

The general solution to these problems is to favor the use of hybrid combinations of mechanical and electronic systems, which have the advantage of dramatically reducing the probability of common mode failures and decreasing sensitivity to external factors. It is this search for the maximization of reliability and ruggedness which is driving the development and application of nanotechnology and MEMS engineering in nuclear weapons science.
To give an important example: modern nuclear weapons use insensitive high-explosives (IHE) which can only be detonated by means of a small charge of sensitive high-explosive that is held out of alignment from the main charge of IHE. Only once the warhead is armed does a MEMS bring the detonator into position with the main charge \cite[p.31]{HUBBE1986-}. Since the insensitive high-explosive in a nuclear weapon is usually broken down into many separate parts that are triggered by individual detonators, the use of MEMS-based detonators incorporating individual locking mechanisms are an important ingredient ensuring the use-control and one-point safety of such weapons \cite{Note-OPS}.

Further improvements on existing nuclear weapons are stemming from the application of nanotechnology to materials engineering. New capacitors, new radiation-resistant integrated circuits, new composite materials capable to withstand high temperatures and accelerations, etc., will enable a further level of miniaturization and a corresponding enhancement of safety and usability of nuclear weapons. Consequently, the military utility and the possibility of forward deployment, as well as the potentiality for new missions, will be increased. 

Consider the concept of a ``low-yield'' earth penetrating warhead \cite{GSPON2003A}. The military appeal of such a  weapon derives from the inherent difficulty of destroying underground targets. Only about 15\% of the energy from a surface explosion is coupled (transferred) into the ground, while shock waves are quickly attenuated when traveling through the ground. Even a few megatons surface burst will not be able to destroy a buried target at a depth or distance more than 100 to 200 meters away from ground zero. A radical alternative, therefore, is to design a warhead which would detonate after penetrating the ground by a few tens of meters or more. Since a free-falling or rocket-driven missile will not penetrate the surface by more than about ten meters, some kind of active penetration mechanism is required. This implies that the nuclear package and its ancillary components will have to survive extreme conditions of stress until the warhead is detonated \cite{Note-EPW}.

\section{Fourth Generation Nuclear Weapons}
\label{fou:0}

First and second generation nuclear weapons are atomic and hydrogen bombs developed during the 1940s and 1950s, while third generation weapons comprise a number of concepts developed between the 1960s and 1980s, e.g., the neutron bomb, which never found a permanent place in the military arsenals. Fourth generation nuclear weapons are new types of nuclear explosives that can be developed in full compliance with the Comprehensive Test Ban Treaty (CTBT) using inertial confinement fusion (ICF) facilities such as the NIF in the US, and other advanced technologies which are under active development in all the major nuclear-weapon states --- and in major industrial powers such as Germany and Japan \cite{GSPON1997-, GSPON2005-}.
 
In a nutshell, the defining technical characteristic of fourth generation nuclear weapons is the triggering --- by some advanced technology such as a superlaser, magnetic compression, nuclear isomers, antimatter \cite{Note-AM}, etc. --- of a relatively small thermonuclear explosion in which a deuterium-tritium mixture is burnt in a device whose weight and size are not much larger than a few kilograms and liters. Since the yield of these warheads could go from a fraction of a ton to many tens of tons of high-explosive equivalent, their delivery by precision-guided munitions or other means will dramatically increase the fire-power of those who possess them --- without crossing the threshold of using kiloton to megaton nuclear weapons, and therefore without breaking the taboo against the first-use of weapons of mass-destruction. Moreover, since these new weapons will use no (or very little) fissionable materials, they will produce virtually no radioactive fallout. Their proponents will define them as ``clean'' nuclear weapons --- and possibly draw a parallel between their battlefield use and the consequences of the expenditure of depleted uranium ammunition \cite{GSPON2002A}.

In practice, since the controlled release of thermonuclear energy in the form of laboratory scale explosions (i.e., equivalent to a few kilograms of high-explosives) at ICF facilities like NIF is likely to succeed in the next 10 to 15 years, the main arms-control question is how to prevent this know-how being used to manufacture fourth generation nuclear weapons. As we have already seen, nanotechnology and micromechanical engineering are integral parts of ICF pellet construction. But this is also the case with ICF drivers and diagnostic devices, and even more so with all the hardware that will have to be miniaturized and `ruggedized' to the extreme in order to produce a compact, robust, and cost-effective weapon.

A thorough discussion of the potential of nanotechnology and microelectromechanical engineering in relation to the emergence of fourth generation nuclear weapons is therefore of the utmost importance. It is likely that this discussion will be difficult, not just because of secrecy and other restrictions, but mainly because the military usefulness and usability of these weapons is likely to remain very high as long as precision-guided delivery systems dominate the battlefield. It is therefore important to realize that the technological hurdles that have to be overcome in order for laboratory scale thermonuclear explosions to be turned into weapons may be the only remaining significant barrier against the introduction and proliferation of fourth generation nuclear weapons. For this reason alone ---and there are many others, beyond the scope of this paper --- very serious consideration should be given to the possibility of promoting an `Inner Space Treaty' to prohibit the military development and application of nanotechnological devices and techniques.

\section*{Acknowledgments}

The author thanks his colleagues at ISRI for their research and comments related to this paper.


\begin{thebibliography}{999}
\bibitem{Note-size} The smaller an electro-mechanical system, the higher its resistance to acceleration. This explains why it is possible to design a shock-proof wrist-watch, while a wall-clock falling on the ground will certainly be damaged. This can easily be verified by calculating the bending of a rod under its own weight, or its resistance to self-traction, which in both cases are proportional to the acceleration and to the length of the rod.  Similar simple arguments show that a small object will cool faster than a larger one, so that under most conditions smaller dimensions are favorable to resist large stresses such as due to intense shocks, extreme vibrations, overheating, fatigue, etc.
\bibitem{Note-EPW}  During free-fall from a height $h \approx 10,000$~meters a weapon of weight $m$ is under the effect of a nearly constant Earth's acceleration $g$ so that its kinetic energy at impact is equal to $mgh$. Similarly, if one assumes as a first approximation that the deceleration $a$ during penetration is constant, the energy required to penetrate a depth $d \approx 10$~meters is given by $mad$.  Equating the two expressions gives then an average deceleration of $a \approx 1,000$~g, which implies that at the moment of impact the deceleration may be as high as 10,000~g.
\end{thebibliography}
\end{document}